\documentclass[a4paper,12pt]{article}
\usepackage{amsmath,amssymb,amsthm}
\usepackage{latexsym,graphicx,color,bbm,subfigure}
\usepackage{pdfsync}   
\usepackage{bbm}

\newcommand{\be}{\begin{equation}}
\newcommand{\ee}{\end{equation}}
\newcommand{\beq}{\begin{eqnarray}}
\newcommand{\eeq}{\end{eqnarray}}

\newcommand{\Tr}{{\rm Tr}}

\newcommand{\bea}{\begin{eqnarray}}
\newcommand{\eea}{\end{eqnarray}}
\def\Tr{ \hbox{\rm Tr}}

\makeatletter \@addtoreset{equation}{section} \makeatother

\begin{document}

\thispagestyle{empty}
\begin{flushright}
IFUP-TH/2013-01
\end{flushright}
\vspace{10mm}
\begin{center}
{\Large   \bf New Confinement Phases from Singular SCFT} 
\\[15mm]
{Simone Giacomelli$^{a,b}$    and Kenichi Konishi$^{c,b}$  
} \footnote{\it e-mail address: si.giacomelli(at)sns.it, ~
konishi(at)df.unipi.it}
\vskip 6mm
 
\bigskip\bigskip
{\it  
$^a$
  Scuola Normale Superiore, 
Piazza dei Cavalieri, 7, Pisa, Italy  
\\
$^b$
  INFN, Sezione di Pisa,
Largo Pontecorvo, 3, Ed. C, 56127 Pisa, Italy 
\\
$^c$  
~Department of Physics ``E. Fermi'', University of Pisa, \\
Largo Pontecorvo, 3, Ed. C, 56127 Pisa, Italy
  }

\vskip 6 mm

\bigskip
\bigskip

{\bf Abstract}\\[5mm]
{\parbox{14cm}{\hspace{5mm}
\small

New types of confining phase emerge when some singular SCFT's  appearing as infrared fixed points
of ${\cal N}=2$ supersymmetric QCD (SQCD) are deformed by an ${\cal N}=1$ adjoint mass term. We make further checks 
on the Gaiotto-Seiberg-Tachikawa (GST) description of these vacua against the symmetry and vacuum counting argument, and show that the GST variables correctly
describe these systems, brought into confinement phase  by the $\mathcal{N}=1$ perturbation. 
Several examples of such vacua,  $USp(2N)$ and $SU(N)$ theories with four flavors and $SO(N)$ theories
with one or two flavors, are  discussed. 
}
}
\end{center}
\newpage
\pagenumbering{arabic}
\setcounter{page}{1}
\setcounter{footnote}{0}
\renewcommand{\thefootnote}{\arabic{footnote}}

%\tableofcontents

\section{Introduction} 

    A proper understanding of a conformal theory appearing as an infrared fixed-point is in many cases of physics of fundamental importance, as it reveals the collective behavior 
of the underlying degrees of freedom, which determines the long-distance physics of the system. Critical phenomena and phase transitions are typical situations in which 
such a consideration plays the central role.  A closely related problem, of particular interest for us, is quark confinement.    Even though the UV behavior of the quark and gluon degrees of freedom is well understood (asymptotic freedom), the collective behavior of color in the infrared is still covered in mysteries. The idea that at certain mass scale the system dynamically Abelianizes and produces Abelian monopoles of the dual $U(1)^{2}$ theory,  and that its dynamical Higgsing induces confinement \cite{TH}, is yet to be demonstrated.  An interesting alternative possibility is that the system does not completely Abelianize, with non-Abelian monopoles of the gauge symmetry breaking
 \beq    {SU(3)   }/   SU(2) \times U(1)   \label{partial}
\eeq
acting as the effective dual degrees of freedom.  As the $u$ and $d$ quarks are light, it is possible that the QCD vacuum is in an $SU(2)$ color-flavor locked phase, in which confinement and chiral symmetry breaking occur simultaneously, via  the condensation of the  non-Abelian monopoles carrying $u$, $d$ flavor charges  \cite{KK}. 
This would alleviate the problem associated with the dynamical Abelianization:  the problem of too-many Nambu-Goldstone bosons
and of the doubling of the meson spectrum.  

At the same time, however, it introduces a new difficulty. In contrast to what happens in the $r$-vacua of the softly broken ${\cal N}=2$ 
supersymmetric QCD,  it is likely that the interactions among non-Abelian monopoles associated with the partial gauge symmetry breaking (\ref{partial}) are asymptotically free and become strong at low energies,  as the sign flip of the beta function is rather difficult in non-supersymmetric QCD.

It is possible that ultimately one must accept the idea that the color magnetic degrees of freedom of QCD are strongly coupled and that confinement and dynamical chiral symmetry breaking are described in a way subtler than a straightforward dual superconductivity picture.  

From this point of view a development of considerable interest is the recent elucidation of the nature of certain infrared-fixed-point SCFT of highest criticality \cite{GST, Simone} in ${\cal N}=2$ supersymmetric QCD with $SU(N)$, $USp(2N)$ and $SO(N)$ gauge groups.  These SCFT's occur at particular points of the vacuum moduli and/or for special choices of the bare quark mass parameters.  A straightforward interpretation of the points of the highest criticality would involve
 monopoles and dyons in an infinite-coupling regime, therefore making their physical interpretation a highly nontrivial task. 
 
 Gaiotto, Seiberg and Tachikawa \cite{GST} applied    the elegant S-dual description discovered by  Argyres and Seiberg \cite{AS}  
for some ``infinitely strongly coupled'' SCFT's   to those SCFT's  appearing as infrared fixed points of  ${\cal N}=2$  $SU(N)$  SQCD with even number of flavors,   solving certain puzzles which remained in earlier studies.  Their analysis has subsequently been generalized by one of us \cite{Simone} to more general class of gauge theories such as $USp(2N)$ and $SO(N)$ with various $N_{f}$. These developments enable us to study new types of confining systems arising from the deformation of these strongly critical SCFT's.  A few results of such a study have recently been presented in \cite{noi2}.  

The purpose of the present paper is to elaborate  more extensively on the properties of these systems.  In particular, we show how these new developments solve the questions left unanswered in some earlier analysis of these singular vacua \cite{CKM,CKKM}, and allow us to understand the way confinement and dynamical symmetry breaking are realized  there. 
In Sections~\ref{uspsection}, \ref{collidingrvacua}, \ref{singularsu(4)}, and  \ref{SO(N)} we study several concrete examples of systems of this kind 
with  $USp(2N)$, $SU(3)$, $SU(4)$, and $SO(N)$ gauge groups and with some special values of $N_{f}$.  We conclude with a brief discussion in Section~\ref{Conclude}.

\section{$USp(2N)$, $N_{f}=4$\label{uspsection}}

The first example we consider is the ${\cal N}=2$ supersymmetric  $USp(2N)$ theory with $N_{f}=2n$ matter hypermultiplets,  perturbed by a small  adjoint scalar mass term, $\mu \Tr \Phi^{2}$.   In the massless limit ($m_{i}=0$, $i=1,\ldots, 2n$)   the theory which survives\footnote{The other set of vacua at the special  point of the Coulomb moduli space  which exist for $N_{f}> N+2$  are not confining \cite{CKKM} and will not be considered here.} is an interacting SCFT with global symmetry  $SO(2N_{f})$,  
first discussed by Eguchi et. al. \cite{EHIY}.  This theory is described by a singular Seiberg-Witten curve,
\begin{equation}
      x y^{2} \sim  \left[ x^{n}(x-\phi_{n}^{2}) \right]^{2}
          - 4\Lambda^{4} x^{2n}
          = x^{2n}(x-\phi_{n}^{2}-2\Lambda^{2})
                  (x-\phi_{n}^{2}+2\Lambda^{2})\;,    \label{Tcheb}
\end{equation}
at  $\phi_{n}^{2}=\pm   2\Lambda^{2}$, that is  $y^{2} \sim  x^{2n}$.  
 The theory at either of these two Chebyshev vacua \footnote{We called these vacua this way  \cite{CKM} as the remaining $N+1-n$  finite vacuum moduli  can be determined by use of a Chebyshev polynomial  \`a  la Douglas-Shenker \cite{DS}.  The first $n-1$ $\phi_{a}$'s have been set to $0$  in (\ref{Tcheb}). } is not  described with a local Lagrangian, as relatively nonlocal massless fields appear simultaneously. 
 
The strategy adopted in \cite{CKM} was to ``resolve'' this vacuum, by introducing generic, nearly equal quark masses $m_{i}$ alongside the adjoint scalar mass $\mu$.   
By requiring the factorization property of the Seiberg-Witten curve to be of maximally Abelian type (the criterion for ${\cal N}=1$ supersymmetric vacua), 
this point was found to split into  various $r$ vacua  which are local $SU(r)\times U(1)^{N-r}$ gauge theories, identical to those appearing in the infrared limit of   $SU(N)$  SQCD (the universality of the infrared fixed points).
\be    \binom{N_{f}}{0} + \binom{N_{f}}{2} + \ldots \binom{N_{f}}{N_{f}} =  2^{N_{f}-1}    \label{one}
\ee
whereas the other vacuum splits into odd $r$ vacua, with the total multiplicity
\be    \binom{N_{f}}{1} + \binom{N_{f}}{3} + \ldots \binom{N_{f}}{N_{f}-1} =  2^{N_{f}-1}\;.  \label{two}
\ee
Due to the exact  ${\mathbbm Z}_{2N+2-N_{f}}$ symmetry of the massless theory,  the singular (EHIY) point actually appears $2N+2-N_{f}$ times, and the 
number of the vacua for generic $\mu, m_{i}$ is given \footnote{For even $N_{f}$ we are considering, the $(N+1-N_{f}/2)$-th element of   ${\mathbbm Z}_{2N+2-N_{f}}$  
exchanges the two Chebyshev vacua \cite{CKM}, so that the number of the vacua is  $ (2N+2-N_{f})\, 2^{N_{f}-1}$  and not  $ (2N+2-N_{f})\, 2^{N_{f}}$. }  by  $ (2N+2-N_{f})\, 2^{N_{f}-1}$.     

A recent study by one of us \cite{Simone}, following the ideas of  \cite{GST}, has shown that this SCFT can be identified  by introducing two different  scaling laws  for the scalar VEVs $u_{i}\equiv \langle \Phi^{i} \rangle$ (the Coulomb branch coordinates)
around the singular point:      
\be  u_{i}\sim\epsilon_{B}^{2i},  \quad ( i=1,\dots,N-n+2);  \qquad u_{N-n+2+i} \sim\epsilon_{A}^{2+2i}, \quad (i=0,\dots,n-2),
\ee
($N_{f}=2n$)  such that  $\epsilon_{B}^{2N+4-2n}=\epsilon_{A}^2;$  ~$\epsilon_{A} \ll \epsilon_{B}$.      
The branch points of the Seiberg-Witten curve separate into various groups of different orders of magnitude and lead to the  infrared physics  described by \cite{Simone} : 
\begin{description}
\item [(i)] $U(1)^{N-n}$ Abelian sector, with massless particles charged under each $U(1)$ subgroup.
\item [(ii)]  The  (in general, non-Lagrangian) A sector  with global symmetry $SU(2)\times SO(4n)$.
\item [(iii)] The B sector which  is free and describes a doublet of hypermultiplets. The flavor symmetry of this system is $SU(2)$.  In contrast to the $SU(N)$ cases studied in \cite{GST}, 
the Coulomb moduli coordinate now includes  $u_{1}$.  We interpret this as representing a low energy effective $U(1)$ gauge field coupled to this hypermultiplet.
\item [(iv)] $SU(2)$ gauge fields coupled weakly to the $SU(2)$ flavor symmetry of the last two sectors.
\end{description} 
For general $N_{f}$  these still involve non-Lagrangian SCFT theory (A sector), and it is not obvious how the $\mu\, \Tr\, \Phi^{2}$ deformation affects the system.
 In a particular case $n=2$  ($USp(2N)$ theory with $N_{f}=4$), however,  the A sector turns out to describe four free doublets 
 of $SU(2)$.  This system  may be symbolically represented, ignoring the  $U(1)^{N-n}$ Abelian sector which is trivial,   as  
 \beq       {\boxed 1} -   SU(2) - {\boxed 4}\;.
 \eeq
  The effect of $\mu \, \Phi^{2}$ deformation of this particular theory can then be analyzed straightforwardly by using the superpotential  
\beq\label{vacmass}   \sqrt{2} \, Q_{0} A_{D} {\tilde Q}^{0} +  \sqrt{2} \, Q_{0} \phi {\tilde Q}^{0} + \sum_{i=1}^{4}  \sqrt{2} \, Q_{i} \phi {\tilde Q}^{i}  +  \mu A_{D} \Lambda  +  {\mu}\,  \Tr \phi^{2}
+  \sum_{i=1}^{4}   m_{i}\, Q_{i} {\tilde Q}^{i}\;,    
\eeq
For equal and nonvanishing masses the system has $SU(4)\times U(1)$ flavor symmetry. In the massless limit the symmetry gets enhanced to $SO(8)$,
in accordance with the symmetry of the underlying $USp(2N)$  theory.    

%In all other cases, e.g.,  two equal masses and two others non equal masses,  the original symmetry is correctly realized by this 
%Lagrangian. 

The vacuum equations are:   
\beq    \sqrt{2} \, Q_{0} {\tilde Q}_{0} + \mu \Lambda =0\;;   \label{eq1}
\eeq
\beq   (\sqrt{2} \, \phi + A_{D})  {\tilde Q}_{0}= Q_{0} \, (\sqrt{2} \, \phi + A_{D}) =0\;;  \label{eq2}
\eeq
\beq  \sqrt{2} \,\,\left[\, \frac{1}{2}  \sum_{i=1}^{4}  Q_{i}^{a}   {\tilde Q}_{b}^{i} -  \frac{1}{4}  \delta_{b}^{a}  Q_{i}{\tilde Q}^{i} +  \frac{1}{2} Q_{0}^{a}{\tilde Q}^{0}_{b}- \frac{1}{4} \delta^{a}_{b}  Q_{0}{\tilde Q}^{0} \, \right] + \mu \, \phi^{a}_{b}=0\;;  \label{eq3}
\eeq
\beq    (\sqrt{2} \,\phi +m_{i} )\, {\tilde Q}^{i}=  Q_{i}\, (\sqrt{2} \,\phi + m_{i})  =0, \qquad \forall i\;.   \label{eq4}
\eeq
The first tells that $Q_{0}\ne 0$.  By gauge choice
\beq  Q_{0} =  {\tilde Q}_{0}= \left(\begin{array}{c}   2^{-1/4}\sqrt{-\mu \Lambda}  \\ 0 \end{array}\right)
\eeq
so that 
\beq    \frac{1}{2} Q_{0}^{a}{\tilde Q}^{0}_{b}- \frac{1}{4}  (Q_{0}{\tilde Q}^{0}) \, \delta^{a}_{b} = \frac{(-\mu \Lambda)}{4 \sqrt{2}} \, \tau^{3}\;.  
\eeq
The second equation can be satisfied by adjusting $A_{D}$.

As in the $m_{i}=0$  case  discussed in  \cite{noi2} (see also \cite{CKM})  we must discard  the  solution 
\beq  \phi=  a\, \tau^{3}, \qquad      a=  \frac{\Lambda}{4}, \qquad Q_{i} ={\tilde Q}_{i}=0, \quad \forall i\;
\eeq
as it involves a fluctuation ($\sim\Lambda$) far beyond the validity of the effective action:  it  is an artefact of the low-energy action.

The true solutions can be found by having one of $Q_{i}$'s canceling the contributions of   $Q_{0}$ and $\phi$ in   Eq~(\ref{eq3}).  Which of $Q_{i}$ is nonvanishing is related to the
value of $\phi$  through Eq~(\ref{eq4}).   For instance, four  solutions can be found by choosing  ($i=1,2,3,4$)
\beq   a=  -\frac{m_{i}}{\sqrt{2}}, \qquad   Q_{i}= {\tilde Q}_{i} = \left(\begin{array}{c}f_i \\0\end{array}\right) ;\qquad  Q_{j}= {\tilde Q}_{j}=0, \quad  j \ne i\;, 
  \label{sol11}  \eeq
such that  
\beq   f_{i}^{2}=    \frac{\mu \Lambda - 4\, a }{\sqrt{2}} =   \mu ( \frac{\Lambda}{\sqrt{2}} + 2 m_{i})\;.  
\eeq

There are four more solutions of the form,  ($i=1,2,3,4$)
\beq   a= + 
\frac{m_{i}}{\sqrt{2}}, \qquad   Q_{i}= {\tilde Q}_{ i} = \left(\begin{array}{c}0 \\g_i\end{array}\right);\qquad  Q_{j}= {\tilde Q}_{j}=0, \quad  j \ne i\;,   
\label{sol22}   \eeq
and 
\beq   g_{i}^{2}=    \frac{-\mu \Lambda + 4\, a }{\sqrt{2}} = -   \mu (\frac{ \Lambda}{\sqrt{2}}  -  2 m_{i})\;.   
  \eeq
Note that the solutions (\ref{sol11}) and (\ref{sol22}) are unrelated to each other  by any $SU(2)$ gauge transformation. 
In all, we have found $2^{3}=8$ solutions consistently with  Eq.~(\ref{one}).

Approaching the equal mass limit the $8$ solutions group into two  set of four nearby vacua, obviously connected by the $SU(4)$. 
So these look like the $4+4 =8$,      two  $r=1$ vacua,     from one of the Chebyshev vacua, see Eq.~(\ref{two}).  The other Chebyshev vacuum should give $1 + 6 + 1=8$ vacua, corresponding to  $r=0,2$ vacua. Where are they?

A possible solution is that in the other Chebyshev vacuum  the superpotential has a similar form as (\ref{vacmass}) but with $Q_{i}$'s carrying different flavor charges.   The $SU(4)$ symmetry of the equal mass theory may be represented as  $SO(6)$:
\beq   \label{vacmassBis}   \sqrt{2} \,Q_{0} A_{D} {\tilde Q}^{0} +  \sqrt{2} \, Q_{0} \phi {\tilde Q}^{0} + \sum_{i=1}^{4}  \sqrt{2} \,Q_{i} \phi {\tilde Q}^{i}  +  \mu A_{D} \Lambda  +  {\mu}\,  \Tr \phi^{2}
+     \sum_{i=1}^{4}   {\tilde m}_{i}\, Q_{i} {\tilde Q}^{i}\;,  
\eeq
where      
\bea   {\tilde m}_{1} =    \frac{1}{4}  (m_{1}+m_{2}- m_{3}-m_{4})\;;  \nonumber \\   
{\tilde m}_{2}=    \frac{1}{4}  (m_{1}-m_{2}+ m_{3}-m_{4})\;;  \nonumber \\ 
  {\tilde m}_{3}=    \frac{1}{4}  (m_{1}-m_{2}- m_{3}+m_{4})\;;\nonumber \\ 
    {\tilde m}_{4}=  \frac{1}{4}  (m_{1}+m_{2} + m_{3} + m_{4})\;.    \label{masses}
\eea
The correct realization of the underlying symmetry in various cases is not obvious, so let us check them  all.  
\begin{description}
\item[(i)]  In the equal mass limit, $m_{i}=m_{0}$,  
\beq       {\tilde m}_{4}=m_{0}, \qquad    {\tilde m}_{2}= {\tilde m}_{3}= {\tilde m}_{4}=0\;, 
\eeq
so  the symmetry is
\beq   U(1)\times SO(6) = U(1)\times SU(4)\;.
\eeq
  Clearly in the $m_{i}=0$ limit the symmetry is enhanced to $SO(8)$. 
  \item[(ii)]   $m_{1}=m_{2}$,    $m_{3}$, $m_{4}$ generic.  
  In this case    ${\tilde m}_{2}= -  {\tilde m}_{3}$ and  ${\tilde m}_{4}$ and  ${\tilde m}_{1}$ are generic, so  the symmetry is   $U(1)\times U(1) \times U(2)$, as in the underlying theory;
  \item[(iii)] $m_{1}=m_{2} \ne 0$,    $m_{3}=m_{4}=0$.    In this case,  ${\tilde m}_{4}=  {\tilde m}_{1}\ne 0$ and ${\tilde m}_{2}=  {\tilde m}_{3} = 0$, so
  obviously the symmetry is $U(2)\times SO(4)$ both in the UV and in  (\ref{vacmassBis}).  
  
  \item[(iv)] $m_{1}=m_{2}=m_{3}\ne 0$,   $m_{4}$ generic.  In this case,  ${\tilde m}_{1}=  {\tilde m}_{2} =  -  {\tilde m}_{3}  \ne 0$,     ${\tilde m}_{4}$ generic. 
  Again the symmetry is $U(3)\times U(1)$ both at the UV and IR.

  \item[(v)]  $m_{1}=m_{2} \ne 0$  and   $m_{3}=m_{4} \ne 0$ but $m_{1}\ne m_{3}$.  In this case  $ {\tilde m}_{2} =  {\tilde m}_{3} = 0$ and  
  $ {\tilde m}_{4}$  and  ${\tilde m}_{1}$ generic.  The flavor symmetry is 
  \beq    SO(4) \times  U(1) \times U(1) =  SU(2)\times SU(2) \times  U(1) \times U(1)\;; 
  \eeq
  this is equal to the symmetry 
  \beq    (SU(2) \times U(1)) \times   (SU(2)\times U(1))
  \eeq
  of the underlying theory.
  \item[(vi)]    $m_{1}\ne 0$,  $m_{2}=m_{3}=m_{4}=0.$  In this case  ${\tilde m}_{1}={\tilde m}_{2}={\tilde m}_{3}={\tilde m}_{4}\ne 0$.
  The symmetry is $U(1)\times SO(6)$ in the UV, and  $U(4)$ in the   infrared. 
   \item[(vii)]    $m_{1}\ne 0$,  $m_{2}\ne 0,$  $ m_{1} \ne m_{2}$,  $m_{3}=m_{4}=0.$  In this case  ${\tilde m}_{1}={\tilde m}_{4}$   ${\tilde m}_{2}={\tilde m}_{3}\ne   {\tilde m}_{1}$.
  The symmetry is $U(1)^{2}\times SO(4)$ in the UV, and  $U(2)\times U(2)$ in the   infrared. 

\end{description}
The cases of masses equal except sign, e.g.,   $m_{1}= - m_{2}$,  are similar. 

Thus in all cases Eq.(\ref{vacmassBis}) has the correct symmetry properties as the underlying theory.  The vacuum solutions which follow from  it are similar to those
found from   Eq.(\ref{vacmass}),  with simple replacement, 
\beq   m_{i} \to {\tilde m}_{i}
\eeq
so  there are  $8$ of them.   The interpretation and their positions in the quantum moduli space (QMS) are different, however.    In the equal mass limit, $m_{i}\to m_{0}$,   The two solutions with 
\beq  a = - \frac{{\tilde m}_{4}}{\sqrt{2}}, \qquad {\rm or} \qquad    a =  \frac{{\tilde m}_{4}}{\sqrt{2}}, 
\eeq
can be regarded as  two  $r=0$ vacua. Note that as $|f_{1}| \ne |g_{1}|$ they correspond to distinct points of the moduli space.  
On the other hand, in the other six vacua  $a=0$ always and     $|f_{i}| = |g_{i}| $, these six solutions correspond to the same point of  
the moduli space: they may be associated with the $r=2$ (sextet) vacua.

\noindent{\bf Remarks:} ~~ The assumption that the $Q_{i}$ fields have different mass assignment in the two Chebyshev vacua, 
as in Eq.~(\ref{vacmass}) and Eq.~(\ref{vacmassBis}) (with (\ref{masses})), is indeed mainly motivated by the fact that the two 
Chebyshev points in QMS are known to behave differently under the mass perturbation \cite{CKM}.  One of the points splits into 
(for nearly equal masses $m_{i}\ne 0$) two nearby groups of $4+4$ vacua  (see Eq.~(\ref{two})) corresponding to two $r=1$ vacua, 
whereas the other is resolved into three groups of $1+1+6$ vacua, which correspond to two $r=0$ and one $r=2$ vacua 
(Eq.~(\ref{one})).  The analysis of this section shows that these properties are precisely reproduced
by our low energy effective action.  
The flavor charges (\ref{masses}) suggest that $Q$'s  are really non-Abelian magnetic monopoles, as semiclassically magnetic 
monopoles appear in the spinor representations of $SO(2N_{f})$.

\section{Colliding $r$ vacua  of the  $SU(3)$, $N_{f}=4$ Theory   \label{collidingrvacua}}  

In the case of $SU(3)$ theory,  with the special value of the number of flavor, $N_{f}=4$,   the effective GST dual is made of 
the A sector describing the three doublets of free hypermultiplets and the B sector,  which is   the most singular SCFT of the $SU(2)$, $N_{f}=2 $ theory  \cite{Argyres:1995xn}     
\beq     D_{3} - SU(2) -  {\boxed 3}  \;.         \label{GSTD3}
\eeq
In order to see the effect of the ${\cal N}=1$ perturbation ${\mu \Phi^{2}}$   in this vacuum,  let us replace the B sector (the $D_{3}$ theory \footnote{We follow the terminology introduced in \cite{Tachi}.}) by  a new $SU(2)$ theory and a bifundamental field $P$,
\beq     SU(2) \, {\stackrel {P}{-}} \,  SU(2) -  {\boxed 3}  \;.    \label{replace}
\eeq
The superpotential has the form, 
\beq       \sum_{i=1}^{3} \sqrt{2}  Q_{i}  \Phi {\tilde Q}^{i}  + \sum_{i=1}^{3}  {\tilde m}_{i}  \,  Q_{i} {\tilde Q}^{i} +      \mu \Phi^{2} +  \sqrt{2}   P  \Phi {\tilde P}
+   \sqrt{2}  {\tilde P} {\chi}     P +  \mu  \, \chi^{2}  +   m^{'} \,  {\tilde P}   P  ,    
\eeq
where $P=P_{a}^{\alpha}$ and $\chi$ is the adjoint scalar of the new $SU(2)$ gauge multiplet.   The new $SU(2)$ intereactions are  asymptotically free and become strong in the infrared. 
As the  $SU(2)$  of  GST is weakly coupled,  the dynamics of the new $SU(2) $ is not affected by it.  
Let us recall that the $D_{3}$ singular SCFT arises in the $SU(2)$ theory with two flavors (with the same bare mass $m^{'}$) as 
the result of collision of a doublet singularity (at $u = m^{' \, 2}$)  with another, singlet vacuum. This occurs when $m^{'}$ coincides 
with the dynamical scale $\Lambda^{'}$.

If we perturb the $D_{3}$ singularity, setting  
\beq  m'  \simeq  \pm \Lambda^{'},  \eeq
but not exactly, the Argyres-Douglas (AD) point \cite{AD} splits as mentioned before in two vacua. Let us analize the resulting systems. The physics of the doublet singularity can be described as follows.  The $P$ system dynamically Abelianizes and gives rise to a superpotential, 
\beq       \sum_{i=1}^{3} \sqrt{2}  Q_{i}  \Phi {\tilde Q}^{i}  + \sum_{i=1}^{3}  {\tilde m}_{i}  \,  Q_{i} {\tilde Q}^{i}+   \mu \Phi^{2}   +  \sqrt{2}   M  \Phi {\tilde M}
+   \sqrt{2}  {\tilde M}  A_{\chi}   M  +  \mu  \, A_\chi  \Lambda^{'},    \label{su3eff}
\eeq
where a doublet of  $M$ represent light Abelian monopoles.  The mass parameters are assumed to have the form \cite{KT}
\bea  {\tilde m}_{1}&=&    \frac{1}{4}  (m_{1}+m_{2}- m_{3}-m_{4})\;;  \nonumber \\
   {\tilde m}_{2}&=&    \frac{1}{4}  (m_{1}-m_{2}+ m_{3}-m_{4})\;;  \nonumber \\
   {\tilde m}_{3} &=&    \frac{1}{4}  (m_{1}-m_{2}- m_{3}+m_{4})\;,      \label{massessu4}  \eea
as in (\ref{masses}) but without ${\tilde m}_{4}$, in terms of the bare quark masses of the underlying $SU(3)$ theory.

Again the flavor symmetry in various cases works out correctly (in all cases a $U(1)$ in the infrared comes from $M$): 
\begin{description}
\item[(i)]  In the equal mass limit, $m_{i}=m_{0}$   ($i=1,2,3, 4$)\;,
\beq          {\tilde m}_{1}= {\tilde m}_{2}= {\tilde m}_{3}=0\;, 
\eeq
so  the symmetry is
\beq   U(1)\times SO(6) = U(1)\times SU(4)\;. 
\eeq
 Note that in contrast to the $USp(2N)$ theory, the symmetry is not enhanced and remains to be $SU(4)\times U(1)$  in the $m_{i}=0$ (so  ${\tilde m}_{i}=0$)  limit 
  \item[(ii)]   $m_{1}=m_{2}$,    $m_{3}$, $m_{4}$ generic.  
  In this case    ${\tilde m}_{2}= -  {\tilde m}_{3}$ and    ${\tilde m}_{1}$ is  generic, so  the symmetry is   $U(1)\times U(1) \times SU(2)$, as in the underlying theory;
  \item[(iii)] $m_{1}=m_{2} \ne 0$,    $m_{3}=m_{4}=0$.    In this case,  $  {\tilde m}_{1}\ne 0$ and ${\tilde m}_{2}=  {\tilde m}_{3} = 0$, so
 the symmetry is $U(2)\times U(2)$  in the UV while  $U(1)\times U(1) \times SO(4)$  in  (\ref{vacmassBis}), which is the same. 
  
  \item[(iv)] $m_{1}=m_{2}=m_{3}\ne 0$,   $m_{4}$ generic.  In this case,  ${\tilde m}_{1}=  {\tilde m}_{2} =  -  {\tilde m}_{3}  \ne 0$. 
  Again the symmetry is $U(3)\times U(1)$ both in the UV and IR.

  \item[(v)]  $m_{1}=m_{2} \ne 0$  and   $m_{3}=m_{4} \ne 0$ but $m_{1}\ne m_{3}$.  In this case  $ {\tilde m}_{2} =  {\tilde m}_{3} = 0$ and  
  $ {\tilde m}_{4}$  and  ${\tilde m}_{1}$ generic.  The flavor symmetry is 
  \beq    SO(4) \times  U(1) \times U(1) =  SU(2)\times SU(2) \times  U(1) \times U(1)\;; 
  \eeq
  this is equal to the symmetry 
  \beq    SU(2) \times U(1) \times   SU(2)\times U(1)
  \eeq
  of the underlying theory.
  \item[(vi)]    $m_{1}\ne 0$,  $m_{2}=m_{3}=m_{4}=0.$  In this case  ${\tilde m}_{1}={\tilde m}_{2}={\tilde m}_{3}\ne 0$.
  The symmetry is $U(1)\times U(3)$ both in the UV and IR.
   \item[(vii)]    $m_{1}\ne 0$,  $m_{2}\ne 0,$  $ m_{1} \ne m_{2}$,  $m_{3}=m_{4}=0.$  In this case  ${\tilde m}_{1} \ne 0$,    ${\tilde m}_{2}={\tilde m}_{3}\ne   {\tilde m}_{1}$.
  The symmetry is $U(1)^{2}\times U(2)$ in the UV, and  $U(1)\times U(1) \times U(2)$ in the   infrared. 

\end{description}

Note that the flavor symmetry in various cases is not the same in $USp(2N)$  and  $SU(N)$ theories.  When  at least two  masses are zero, the symmetry is larger in the $USp(2N)$ theory, 
so the exact matching of the flavor symmetry in the UV and in the IR   is  quite nontrivial.  

The vacuum equations are now  
\beq    M {\tilde M} + \mu \Lambda^{\prime} =0\;;   \label{eq1bis}
\eeq
\beq   (\phi + A_{\chi})  {\tilde M}=M \,(\phi + A_{\chi})  =0\;;  \label{eq2bis}
\eeq
\beq  \sqrt{2} \,\left[\,  \frac{1}{2}  \sum_{i=1}^{3}  Q_{i}^{a}   {\tilde Q}_{b}^{i} -  \frac{1}{4}  \delta_{b}^{a}  (Q_{i}{\tilde Q}^{i})  +  \frac{1}{2}M^{a}{\tilde M}_{b}- \frac{1}{4} \delta^{a}_{b} M {\tilde M}\,\right]+ \mu \, \phi^{a}_{b}=0\;;  \label{eq3bis}
\eeq
\beq    (\phi + {\tilde m}_{i} )\, {\tilde Q}^{i}=  Q_{i}\, (\phi + {\tilde m}_{i})  =0, \quad \forall i\;.   \label{eq4bis}
\eeq
The first says that $M \ne 0$.  By gauge choice
\beq  M =  {\tilde M}= \left(\begin{array}{c}  2^{-1/4}\sqrt{-\mu \Lambda^{\prime}}  \\ 0 \end{array}\right)   \label{gchoice}
\eeq
so that 
\beq    \frac{1}{2} M^{a}{\tilde M}_{b}- \frac{1}{4}  (M{\tilde M}) \, \delta^{a}_{b} = \frac{(-\mu \Lambda^{'})}{4\sqrt{2}} \, \tau^{3}\;.  
\eeq
The second equation   is satisfied by adjusting $A_{\chi}$, whichever valur $\phi$ takes. The solution of the third equation (\ref{eq3bis}) without $Q$ vevs is not acceptable as it implies a
large ($O(\Lambda')$)  vev for $\phi.$   We are led to conclude that  ($i=1,2,3$): 
\beq   a=  -{\tilde m}_{i}, \qquad   Q_{i}= {\tilde Q}_{i} = \left(\begin{array}{c}h_i \\0\end{array}\right) ;\qquad  Q_{j}= {\tilde Q}_{j}=0, \quad  j \ne i \;.  
\eeq
such that  
\beq   h_{i}^{2}=    \frac{\mu \Lambda'}{\sqrt{2}}   +  2\, {\tilde m}_{i} \, \mu 
\eeq
or 
\beq   a=  {\tilde m}_{i}, \qquad   Q_{i}= {\tilde Q}_{ i} = \left(\begin{array}{c}0 \\  k_i\end{array}\right);\qquad  Q_{j}= {\tilde Q}_{j}=0, \quad  j \ne i  \;,   
\label{sol111}
\eeq
and 
\beq   k_{i}^{2}=   - \frac{\mu \Lambda' }{\sqrt{2}}  +   2\, {\tilde m}_{i} \, \mu \;.    
\label{sol222}   \eeq
These  give six vacua (corresponding to the $r=2$ vacua).  Where are other, $r=0,1$ vacua?

Now in the singlet vacuum the low energy physics of the new $SU(2)$ theory is
an Abelian gauge theory with a single monopole, $N$,  thus our effective superpotential is similar with (\ref{su3eff}) but with $N$ field having no coupling to the weak GST $SU(2)$
gauge fields: 
\beq       \sum_{i=1}^{3} \sqrt{2}  Q_{i}  \Phi {\tilde Q}^{i}  + \sum_{i=1}^{3}  {\tilde m}_{i}  \,  Q_{i} {\tilde Q}^{i}+   \mu \Phi^{2}   
+   \sqrt{2}  {\tilde N}  A   N  +  \mu  \, A  \Lambda^{'}+   m^{'} \,  {\tilde N}  N.   \label{su3effBis}
\eeq
Now the $U(1)$ part gets higgsed as usual,  and the GST $SU(2)$ gauge theory becomes asymptotically free, having ${\tilde N}_{f}=3$  hypermultiplets with small masses ${\tilde m}_{i}$. 
The infrared limit of this theory is well known: there is one vacuum with four nearby singularities and one singlet vacuum \cite{SW2}. In the quadruple vacuum,  the mass perturbation  ${\tilde m}_{i}$
give  four nearby vacua, the light hypermultiplets have masses, in the respective vacua \cite{KT},
\beq   {\hat m}_{1}=   {\tilde m}_{1}+  {\tilde m}_{2}+ {\tilde m}_{3} = \frac{1}{4} ( 3 m_{1}- m_{2}-m_{3}- m_{4})\;;
\eeq
\beq   {\hat m}_{2}=   -{\tilde m}_{1}-  {\tilde m}_{2}+ {\tilde m}_{3} = \frac{1}{4} ( 3 m_{4}- m_{1}-m_{2}- m_{3})\;;
\eeq
\beq   {\hat m}_{3}=  - {\tilde m}_{1}+  {\tilde m}_{2} -{\tilde m}_{3} = \frac{1}{4} ( 3 m_{3}- m_{1}-m_{2}- m_{4})\;;
\eeq
\beq   {\hat m}_{4}=   {\tilde m}_{1} -  {\tilde m}_{2} - {\tilde m}_{3} = \frac{1}{4} ( 3 m_{2}- m_{1}-m_{3}- m_{4})\;:
\eeq
they correctly represent the physics of the $r=1$ vacuum where the light hypermultiplets appear in the ${\underline 4}$ of the underlying $SU(N_{f})= SU(4)$ group.

Finally, in the singlet vacuum of the new $SU(2)$,  ${\tilde N}_{f}=3$ theory, the light hypermultiplet is a singlet of the flavor group, so it is also a singlet of the original 
$SU(4)$ flavor group.  

%%%%%%%%%%%%
%************
%Now that we have ascertained that the vacuum structure is correctly reproduced by the  system (\ref{replace}), we can go to the massless limit $m_{i} \to 0$.
%As the nonvanishing VEV arises only from 

 \section{Singular $r=2$ vacua of the $SU(4)$, $N_{f}=4$ Theory \label{singularsu(4)}} 
 
In the case of the higher singularity of $SU(4)$, $N_{f}=4$ theory discussed in \cite{noi2},  the GST  dual description is  
\beq      D_{4} - SU(2)  -   {\boxed 3}  \label{noncollidr}
\eeq
where $D_{4}$ is the most singular SCFT of  $SU(3)$ theory with ${\tilde N}_{f}=2$ flavors and  $ {\boxed 3} $  represents three doublets of free hypermultiplets.   The two SCFTs are
coupled weakly through the weak  $SU(2)$   interactions \cite{GST}. 
We therefore replace the above with another system 
\beq     SU(3) \, {\stackrel {B}{-}} \,  SU(2) -  {\boxed 3}  
\eeq
with a bifundamental field $B^{\alpha}_{a}$ carrying both $SU(3)$ and $SU(2)$ charges, that is, 
\beq      \sum_{i=1}^{3} \sqrt{2}  Q_{i}  \Phi {\tilde Q}^{i}  + \sum_{i=1}^{3}  {\tilde m}_{i}  \,  Q_{i} {\tilde Q}^{i} +      \mu \Phi^{2} +  \sqrt{2}   B  \Phi {\tilde B}
+   \sqrt{2}  {\tilde B} {\chi}     B   +  \mu  \, \chi^{2}  +   m^{''} \,  {\tilde B}   B\;.    \label{particular}
\eeq
where $\chi$  is the adjoint scalar of the new $SU(3)$ and ${\tilde m}_{i}$ are given by  (\ref{massessu4}). For simplicity we 
have set the mass parameters for $\Phi$ and $\chi$ to be equal.  ${\tilde m}_{i}=0$, $i=1,2,3$ in the equal mass limit of the underlying theory,  $m_{i}=m_{0}$, $i=1,\ldots, 4$,   
so the system has the correct flavor symmetry,  $SO(6)\times U(1)= SU(4)\times U(1)$.   The flavor symmetry in various cases of unequal masses works out as in Subsection~\ref{collidingrvacua}.  

The $SU_{GST}(2)$  interactions are weak and do not affect significantly the $SU(3)$ gauge interactions.  Actually,  in order to study the system (\ref{noncollidr}), we must  focus our attention to one 
particular $SU(3)$  vacuum  (i.e., $D_{4}$ SCFT).   $D_{4}$ SCFT  appears as the $r=\frac{N_{f}}{2} =1$  vacuum of the new 
$SU(3)$, ${\tilde N}_{f}=2$ theory in the limit  $m^{''} \to 0$ (see Appendix D of \cite{noi2}).   In contrast to the case discussed 
in the previous subsection, the $r=1$ vacuum does not collide with the $r=0$ vacuum.

The $SU(3)$, ${\tilde N}_{F}=2$ theory is asymptotically free and becomes strongly coupled
in the infrared. For $m^{''}\neq0$ the low energy dynamics at the $r=1$ vacuum is described by a $U(1)^{2}$ theory \cite{SW2}:  two types of  
massless monopole hypermultiplets 
$M$ and $N$  appear, each carrying one of the local $U(1)$ charges, and one of them  ($M$) is a doublet of the  flavor 
$SU({\tilde N}_{F})=SU_{GST}(2)$.
Therefore the low-energy effective superpotential is given by 
 \bea   &&  \sum_{i=1}^{3} \sqrt{2}  Q_{i}  \Phi {\tilde Q}^{i}  + \sum_{i=1}^{3}  {\tilde m}_{i}  \,  Q_{i} {\tilde Q}^{i} +      \mu \Phi^{2} +  \sqrt{2}   M  \Phi {\tilde M}
+   \sqrt{2}  {\tilde M} A_{\chi}  M  +  \mu  \, A_{\chi}  \Lambda^{'}    +   m^{''} \,  {\tilde M}   M  \nonumber \\
&&  +      \sqrt{2}  {\tilde N} A  N  +  \mu  \, A  \Lambda^{'}    \label{abelian}
\eea
where $M$ is now a doublet of Abelian monopoles and $N$ is the Abelian monopole, singlet of the flavor $SU({\tilde N}_{F})=SU(2)$. The vacuum equations are  
\beq    \sqrt{2} \, M {\tilde M} + \mu \Lambda^{\prime} =0\;;  
\eeq
\beq   ( \sqrt{2} \phi + A_{\chi}+ m^{''}  )  {\tilde M}=M \,( \sqrt{2} \phi + A_{\chi} +  m^{''} )  =0\;;  
\eeq
\beq  \sqrt{2} \left[\, \frac{1}{2}  \sum_{i=1}^{3}  Q_{i}^{a}   {\tilde Q}_{b}^{i} -  \frac{1}{4}  \delta_{b}^{a}  (Q_{i}{\tilde Q}^{i})  +  \frac{1}{2}M^{a}{\tilde M}_{b}- \frac{1}{4} \delta^{a}_{b} M {\tilde M}\,\right]  + \mu \, \phi^{a}_{b}=0\;;  
\eeq
\beq    ( \sqrt{2} \phi + {\tilde m}_{i} )\, {\tilde Q}^{i}=  Q_{i}\, ( \sqrt{2} \phi + {\tilde m}_{i})  =0, \quad \forall i\;.   
\eeq
The solution of these equations are given by Eqs. (\ref{gchoice})-(\ref{sol222})  with the replacement ${\tilde m}_{i} \to m_{i}$. 
We therefore find six vacua, corresponding to the $r=2$  vacua of the underlying theory.

The degeneracy of vacua can also  be determined integrating out the $SU(3)$ $\psi$ field and adding the ADS superpotential. This procedure 
will inevitably produce the whole set of vacua of the model, including all vacua of the $SU(3)$ theory, whereas we are interested only in one of the  $r={\tfrac{{\tilde N}_{f}}{2}}=  1$ vacuum, since we are interested in the $D_4$ sector. Our strategy will be to make the computation in the general case and then discard all the 
unwanted solutions. Integrating out $\psi$ the effective superpotential becomes 
\beq\nonumber 
\mathcal{W}=\sum_{i=1}^{3} \sqrt{2}  Q_{i}  \Phi {\tilde Q}^{i}  + \sum_{i=1}^{3}  {\tilde m}_{i} Q_{i} {\tilde Q}^{i} +\mu \Phi^{2}
+m\Tr M+\frac{\mu^3\Lambda^4}{\det M}+\Tr(\Phi M)-\frac{1}{2\mu}\left(\Tr M^2-\frac{(\Tr M)^2}{3}\right),
\eeq where $M_{ab}$ is the meson field $\tilde{B}_{a\alpha}B_{b}^{\alpha}$. The meson matrix can be supposed diagonal and we 
will parametrize it as $M=aI+2b\tau_3$, so
$$\Tr(\Phi M)=\Phi_3b,\; \det M=a^2-b^2,\;\Tr M=2a,\; \Tr M^2=2(a^2+b^2)\;. $$ 
The superpotential then becomes
\beq
\mathcal{W}=\sum_{i=1}^{3} \sqrt{2}  Q_{i}  \Phi {\tilde Q}^{i}  + \sum_{i=1}^{3}  {\tilde m}_{i} Q_{i} {\tilde Q}^{i} +\mu \Phi^{2}
+\Phi_3b+2ma+\frac{\mu^3\Lambda^4}{a^2-b^2}+\frac{4a^2}{6\mu}-\frac{a^2+b^2}{\mu}\;.
\eeq Modulo a gauge choice we can diagonalize the $\Phi$ field, so that $\Phi_3$ is the only nonvanishing component. We thus find 
the following F-term equations:
\beq2m-\frac{2a}{3\mu}-\frac{2\mu^3\Lambda^4}{(a^2-b^2)^2}a=0\;,\eeq
\beq\Phi_3-\frac{2b}{\mu}+\frac{2\mu^3\Lambda^4}{(a^2-b^2)^2}b=0\;,\eeq
\beq(\sqrt{2}\Phi+m_i)Q_i=0,\quad b+\mu\Phi_3+\frac{\sqrt{2}\tilde{Q^i}Q_i\vert_{3}}{2}=0\;,\eeq
where $\tilde{Q}Q\vert_3$ is the component proportional to $\tau_3$. If the vev of $Q$ is nonzero, then $\Phi_3=\pm\sqrt{2}m_i$ 
(as before, only one $Q_i$ can have vev and one of the two components must vanish). Since there are six possibilities we will get 
the six vacua we were looking for. The last equation then tells that 
$$\tilde{Q^i}Q_i=\pm 2\mu m_i-b\;.$$
 The first two equations imply that $a^2-b^2\propto\mu^2\Lambda^2$.
If the vev of $Q$ is zero we have two possibilities: b nonzero and $\mu\Phi_3=-b$. But then we get $b\sim\mu\Lambda$ which in turn 
implies that $\Phi_3\sim\Lambda$ and this solution should be discarded. The other possibility is 
$b=\Phi_3=0$ and then from the first equation $a\sim\mu\Lambda$. We get four solutions which precisely correspond to the $r=0$ 
vacua of the $SU(3)$ theory. Since we are not interested in these vacua we simply discard them. 

The solutions corresponding to the 
$r=2$ vacua of our theory are correctly characterized by a nonvanishing $Q$ condensate and thus the pattern of flavor symmetry breaking 
is the expected $$U(4)\rightarrow U(2)\times U(2)\;.$$
Indeed we can proceed as before and choosing $m\neq0$ the singular point describing the $D_4$ theory evolves into a $r=1$ vacuum 
whose massless spectrum includes two vector multiplets that we will denote $A$ and $B$, two hypermultiplets charged under e.g.,  $A$ and forming a doublet of the $SU(2)$ flavor symmetry of the theory and a third hypermultiplet charged under $B$ which is 
a singlet of the flavor symmetry. The effective superpotential is thus
\beq\nonumber
\mathcal{W}=\sum_{i=1}^{3} \sqrt{2}  Q_{i}  \Phi {\tilde Q}^{i}  + \sum_{i=1}^{3}  {\tilde m}_{i} Q_{i} {\tilde Q}^{i} +\mu \Phi^{2}
+ \sqrt{2}\tilde{M}{\phi}M+ \sqrt{2}\tilde{M}{A}M+\sqrt{2}\tilde{R}{B}R+\mu_1\Lambda A +\mu_2\Lambda B\;. 
\eeq
This correctly describes the physics of the perturbed $r=2$ vacuum of $SU(4)$ SQCD with four flavors, leading to six vacua. The computation 
proceeds as in the previous sections.

\section{Singular points of $SO(N)$ SQCD \label{SO(N)}}

In \cite{Simone} Chebyshev points of $SO(N)$ theories were analyzed as well. The outcome was the by now familiar two-sector structure (for $SO(2N)$ SQCD 
with even $N_f$ or $SO(2N+1)$ SQCD with $N_f$ odd): one hypermultiplet charged under an Abelian gauge group and a SCFT which can be described in terms of the 6d $D_N$ theory compactified 
on a three-punctured sphere \cite{Tachi}. These two sectors are coupled as before through an infrared free $SU(2)$ vector multiplet. The analysis of the 
$\mathcal{N}=1$ breaking in the general case is still out of reach but we can analyze in detail a couple 
of examples with low number of flavors, since as expected the superconformal sector simplifies enough to make the problem approachable. We will study 
the cases $N_f=1,2$ which already involve a nontrivial structure hard to guess without performing the analysis in the above mentioned paper.

\subsection{$SO(2N+1)$ theory with one flavor}

The SW curve at the Chebyshev point  of $SO(2N+1)$ SQCD  with one flavor becomes $y^2=x^4$. The superconformal sector entering the GST description 
becomes free in this case and describes one hypermultiplet in the adjoint of $SU(2)$, thus saturating its beta function. Notice that starting in the 
UV from a theory with a single matter field in the vector representation, we end up with an infrared effective description involving an $SU(2)$ 
gauge group coupled to matter fields in different representations! The expectation from the semiclassical analysis \cite{CKKM}  is that the $SU(2)$ flavor symmetry 
is dynamically broken to $U(1)$ when the mass term $\mu\Tr\Phi^2$ is turned on. Furthermore, if we give mass to the flavor (or to the hyper in the 
adjoint in the effective description) we expect to get two vacua  ($2^{N_{f}}=2$). Since our infrared effective theory admits a Lagrangian description, all these properties 
should be reproduced by the equations of motion. We will now check that this is case.

In the $\mathcal{N}=1$ language we describe the hypermultiplet in the adjoint using two chiral multiplets $X_1$ and $X_2$. The superpotential is
\beq\label{SO1}\mathcal{W}=\sqrt{2} \tilde{Q} A_{D} Q +  \sqrt{2} \tilde{Q} \Phi Q +  \mu A_{D} \Lambda  +  {\mu}\,  \Tr \Phi^{2}
+ \sqrt{2}i\Tr(\Phi[X_1,X_2])+ m\Tr(X_1X_2)\;.\eeq 
The variation with respect to $A_D$ tells that $Q$ has a nonvanishing vev. By gauge choice we can 
set $Q_2=0$ and then the variation with respect to $\tilde{Q}$ implies that $\Phi$ is diagonal. The variation with respect to the fields in the adjoint 
give the equations (we write them as $X=X^{a}\tau_{a}$):
\beq\label{phi}\sqrt{2} \tilde{Q} \tau_{a} Q + \mu\Phi_{a}+\sqrt{2}i\Tr(\tau_{a}[X_1,X_2])=0\;,\eeq
\beq\label{x1}\frac{m}{2}X_{2}^{a} + \sqrt{2}i\Tr(\Phi[\tau^{a},X_2])=0\;,\eeq
\beq\label{x2}\frac{m}{2}X_{1}^{a} - \sqrt{2}i\Tr(\Phi[\tau^{a},X_1])=0\,.\eeq
Since $\Phi_1=\Phi_2=0$, equations (\ref{x1},\ref{x2})  imply that $X_{1}^{3}=X_{2}^{3}=0$ (it suffices to note that $\Tr(\tau_3[\tau_3,\cdotp])=0$).
The nontrivial equations become then
 $$\mu\Phi_3-\frac{\mu\Lambda}{2}-\frac{\sqrt{2}}{2}\epsilon_{ij3}X_{1}^{i}X_{2}^{j}=0,\quad 
\frac{m}{2}X_{2}^{i}-\frac{\sqrt{2}}{2}\epsilon_{ij3}\Phi_3X_{2}^{j}=0,\quad \frac{m}{2}X_{1}^{i}+\frac{\sqrt{2}}{2}\epsilon_{ij3}\Phi_3X_{1}^{j}=0\;.$$
Notice that the above equations imply that none of the unknowns can vanish (if one of the $X_i$'s vanish we would have $\Phi_3\sim\Lambda$, which we 
must discard as explained in the previous sections). Setting $X_1^{i=1}\equiv a$, $X_1^{i=2}\equiv b$, the second equation leads to the system 
$$\frac{m}{2}a+\frac{\Phi_3}{\sqrt{2}}b=0,\quad \frac{m}{2}b-\frac{\Phi_3}{\sqrt{2}}a=0\;.$$
 Writing $a$ in terms of $b$ and $\Phi_3$ using the first 
relation and substituting in the second we directly get $\Phi_3^2=-m^2/2$ and thus $a=\pm ib$. Using an analogous argument the third equation leads 
to $d=\pm ic$, where $X_2^{i=1}\equiv c$, $X_2^{i=2}\equiv d$. Notice that if we choose e.g. $a=+ib$ we are forced to set $d=+ic$ and not $-ic$, 
otherwise the term $\epsilon_{ij3}X_{1}^{i}X_{2}^{j}$ would vanish. Since the D-term equations imply that $\vert b\vert=\vert c\vert$, we get two solutions 
as expected. In the massless limit $m=0$, both the gauge and flavor symmetries are broken by the vevs of $Q$ and $X_i$'s. However, a diagonal combination 
of the gauge and flavor Abelian subgroups leaves the vevs invariant. We thus recover the expected $U(1)$ flavor symmetry.

\subsection{$SO(2N)$ theory with two flavors  \label{SO(N)two}    }

The SW curve at the Chebyshev points in these theories becomes $y^2=x^6$  \cite{CKKM}. In \cite{Simone} it was found that the superconformal sector does not become 
free in this case but turns out to be a well known Lagrangian SCFT: $SU(2)$ SQCD with four flavors. Symbolically the system can be represented as 
\beq      {\boxed 1}\, {\stackrel {Q}{-}}  \,  SU(2) \,{\stackrel {M_{i}}{-}}  \,SU(2)\;.
\eeq
An $SU(2)$  subgroup of the $SO(8)$ flavor symmetry is gauged 
in the present context, leaving the commutant $USp(4)$ ungauged, matching the UV flavor symmetry of the theory. The $USp(4)$ symmetry tells us that 
the low energy theory at the Chebyshev point is a $SU(2)\times SU(2)$ gauge theory with two hypermultiplets in the bifundamental and one doublet 
charged under only one of the $SU(2)$ factors. The semiclassical analysis predicts that the flavor symmetry is dynamically broken to $U(2)$ when 
we turn on the mass term for the chiral multiplets in the adjoint, while keeping the bifundamentals massless \cite{CKKM}. If we give mass to the bifundamentals as well, 
we expect to find $2^{N_f}=4$ vacua. We shall now see that the equations of motion for the infrared effective theory reproduce all these features.

We indicate in $\mathcal{N}=1$ notation the hypermultiplets in the bifundamental with $M_1$, $\tilde{M}_1$, $M_2$, $\tilde{M}_2$ and the chiral 
multiplets in the adjoint with $\Phi$ and $\Psi$. The superpotential will then be (the sum over $i=1,2$ for the bifundamentals is implied)
\beq\label{so2}\begin{aligned}
\mathcal{W}=&\sqrt{2} \tilde{Q} A_{D} Q +  \sqrt{2} \tilde{Q} \Phi Q  + 
m_i\Tr(\tilde{M}_iM^i)+\sqrt{2}\Tr(\tilde{M}_i\Phi M^i)+\sqrt{2}\Tr(M^i\Psi \tilde{M}_i)\\
&+  \mu A_{D} \Lambda  +  {\mu} \Tr \Phi^{2} + {\nu} \Tr \Psi^{2},\end{aligned}
\eeq where $\mu$ and $\nu$ are of the same order.
As usual, the variation with respect to $A_D$ implies that the doublet $Q$ has non vanishing vev. We can then use the gauge freedom to set $Q_2$ 
to zero and to diagonalize $\Psi$. The equation for $\tilde{Q}$ will then imply that $\Phi$ is diagonal too. The equations coming from the variation 
of $\Phi$ and $\Psi$ are then 
\beq\label{phibis}\mu\phi_3-\frac{ \mu \Lambda}{2}+\sqrt{2}\Tr(\tilde{M}_i\tau_3 M^i)=0,\quad \sqrt{2}\Tr(\tilde{M}_i\tau_{1,2} M^i)=0\;,   \label{this}   \eeq
\beq\label{psibis}\nu\Psi_3+\sqrt{2}\Tr(M^i\tau_3 \tilde{M}_i)=0,\quad \sqrt{2}\Tr(M^i\tau_{1,2} \tilde{M}_i)=0\;.    \label{that}  \eeq 
Since this will play a role later in the derivation, we would like to draw the reader's attention to the fact that we require 
the vevs of both $\Phi$ and $\psi$ to be much smaller than $\Lambda$. This in particular implies that $\Tr(\tilde{M}_i\tau_3 M^i)$ 
and $\Tr(M^i\tau_3 \tilde{M}_i)$ cannot be of the same order (the first should be much larger than the second in order to compensate 
the term proportional to $\mu\Lambda$ in (\ref{phibis})). The variation with respect to 
the bifundamental fields gives $$\sqrt{2}\Phi M_i + m_iM_i + \sqrt{2}M_i\Psi=0,\quad \sqrt{2}\Psi\tilde{M}_i+m_i\tilde{M}_i+\sqrt{2}\tilde{M}_i\Phi=0\;.$$ 
It is convenient to rewrite these two equations in matrix form:
\beq\label{matr}\begin{aligned}
\left(\begin{array}{cc}
(m_i+\frac{\Phi_3}{\sqrt{2}}+\frac{\Psi_3}{\sqrt{2}})a_i & (m_i+\frac{\Phi_3}{\sqrt{2}}-\frac{\Psi_3}{\sqrt{2}})b_i\\
(m_i-\frac{\Phi_3}{\sqrt{2}}+\frac{\Psi_3}{\sqrt{2}})c_i & (m_i-\frac{\Phi_3}{\sqrt{2}}-\frac{\Psi_3}{\sqrt{2}})d_i\\
\end{array}\right)&=0\;,\\
\left(\begin{array}{cc}
(m_i+\frac{\Phi_3}{\sqrt{2}}+\frac{\Psi_3}{\sqrt{2}})e_i & (m_i-\frac{\Phi_3}{\sqrt{2}}+\frac{\Psi_3}{\sqrt{2}})f_i\\
(m_i+\frac{\Phi_3}{\sqrt{2}}-\frac{\Psi_3}{\sqrt{2}})g_i & (m_i-\frac{\Phi_3}{\sqrt{2}}-\frac{\Psi_3}{\sqrt{2}})h_i\\
\end{array}\right)&=0\;,\end{aligned}
\eeq 
where we have set (we will take into account D-terms later) 
$$ M_i=\left(\begin{array}{cc}
a_i & b_i\\
c_i & d_i\\
\end{array}\right),\qquad\tilde{M}_i=\left(\begin{array}{cc}
e_i & f_i\\
g_i & h_i\\
\end{array}\right),\quad i=1,2 \;.$$
It is clear that for $m_1$ and $m_2$ generic, equation (\ref{matr}) requires that some entries of $M_i$ and $\tilde{M}_i$ vanish. 

One can check that there are no solutions if only one of the bifundamentals (for instance,  $M_1$) is to have nonvanishing vev. This can be shown 
as follows: it is easy to see that at least two entries (both for $M_1$ and $\tilde{M}_1$) must be zero. Taking into account 
the rightmost equations in (\ref{phibis}) and (\ref{psibis}), one can easily check that actually at most one entry can be different from 
zero, but then $\Tr(\tilde{M}_i\tau_3 M^i)$ and $\Tr(M^i\tau_3 \tilde{M}_i)$ differ at most by a sign and are thus of the same 
order and this is in conflict with the observation we made before. 

We are then forced to let both $M_1$ and $M_2$ be nontrivial. This can be achieved by imposing e.g. the equations 
\beq\label{solu}m_1+\frac{\Phi_3}{\sqrt{2}}+\frac{\Psi_3}{\sqrt{2}}=0,\qquad m_2+\frac{\Phi_3}{\sqrt{2}}-\frac{\Psi_3}{\sqrt{2}}=0\;,  \label{choice}  \eeq 
which determine both $\Phi_3$ and $\Psi_3$ in terms of the mass parameters.
Clearly both $M_1$ and $M_2$ can have only one nonvanishing  entry.  

Let us now count the number of possible solutions: one naively 
has four possible choices for $M_1$; in two cases the matrix is diagonal and in the other two offdiagonal. Actually, the action 
of  subgroup   
\beq    T =  \left(\begin{array}{cc}0 & 1 \\-1 & 0\end{array}\right) 
\eeq
of the second $SU(2)$ gauge factor interchanges these two 
possibilities and we may assume, e.g., that $M_1$ is diagonal.  There are two solutions according to which diagonal element is chosen to be nonvanishing.  

The other two solutions (in which $M_{1}$ is offdiagonal) are gauge equivalent to these and should not 
be considered distinct \footnote{We cannot act with an analogous subgroup of the other gauge factor as 
it is already broken by the $Q$ vev.}.      Having $M_{1}$ of diagonal form, equation (\ref{matr}) then implies that $M_2$ is offdiagonal and we have two possible 
choices. If we now take into account the D-term condition  we find that (modulo a phase) $\tilde{M}_1=M_1^{\dagger}$ and 
$\tilde{M}_2=M_2^{\dagger}$. Each one of the four possible choices lead then to a single solution once equations (\ref{phibis}), (\ref{psibis}) 
are taken into account. We thus find    four solutions as anticipated. One of the solutions, corresponding to the choice, (\ref{choice}), takes the form, 
\beq   M_{1}=  \left(\begin{array}{cc}  a_{1}  & 0 \\0 & 0 \end{array}\right)\;; \qquad M_{2} = \left(\begin{array}{cc}0 & 0 \\  g_{2}  & 0\end{array}\right)\;,
\eeq
where  $a_{1}$ and $g_{2}$ are determined by Eqs.~(\ref{this}), (\ref{that}), (\ref{choice}) and are of the order of $O(\mu \Lambda, \mu\, m_{i})$.

In the massless limit the $\Phi$ and $\Psi$ vevs go to zero. The $Q$ condensate breaks the first $SU(2)$ gauge symmetry factor and 
the vev of the bifundamentals breaks the second $SU(2)$ gauge factor. The $USp(4)$ flavor symmetry is broken as well;  however, 
there is a diagonal combination of the global $SU(2)$ gauge transformations (coming from the second gauge factor) and (an $SU(2)$ 
subgroup of) flavor transformations which acts trivially on our solution of the field equations and thus remains  unbroken. Furthermore, 
the second Cartan generator of the flavor symmetry group of the theory can combine with the Cartan of the first $SU(2)$ gauge group 
to give the generator of a $U(1)$ group which is unbroken. The color-flavor locking mechanism thus leads to the 
$U(2)$  unbroken global symmetry,   which is the correct unbroken symmetry expected from the analysis made at large $\mu$ \cite{CKKM}.

\section{Discussion \label{Conclude}} 

The fate of an ${\cal N}=2$ SCFT  upon deformation by  ${\cal N}=1$, adjoint mass perturbation, $\mu  \Phi^{2}$, can be 
of  several different types.   A nontrivial ${\cal N}=2$ SCFT  in the UV  might smoothly flow into an ${\cal N}=1$ SCFT in the infrared (see \cite{TachiWecht} for some
 beautiful observations).   An infrared fixed-point  SCFT in an ${\cal N}=2$ theory might get lifted upon $\mu  \Phi^{2}$ deformation,
  as in the case of the original AD point in the pure  ${\cal N}=2$, $SU(3)$ theory. 
  
 Infrared fixed-point ${\cal N}=2$ SCFT's might also be brought into confinement phase, as shown in the original  Seiberg-Witten work \cite{SW1,SW2}, in 
 the case of local $r$ vacua \cite{CKM}, or  in the cases of singular SCFT's   discussed in the present paper.    What distinguishes these systems is the presence of $U(1)$ factors in the effective gauge symmetry.  More precisely the property required is the nontrivial fundamental group, 
 \beq   \pi_{1} (G_{eff})  \ne {\bf 1}\;
 \eeq
 where $G_{eff}$ is the low-energy gauge group,  and that  all the $U(1)$ factors are broken upon $\mu \, \Phi^{2}$ perturbation. If the underlying gauge group is simply connected, the 
 vortices of the low-energy theory should not exist in the full theory. If the low-energy theory is magnetic, then the condensation leading to the breaking
 $G_{eff} \to {\bf 1}$  implies confinement of color. 
 
 What was not known in earlier studies  \cite{CKM,CKKM,Konishi:2005qt}  is what happens in the singular Chebyshev vacua 
 (EHIY points), and if the system would  be brought into confinement phase, which kind of confinement phase it would be.  Since such a  system apparently involved
 (infinitely) strongly-coupled, relatively nonlocal monopoles and dyons,  it was not at all evident whether or not the standard (weakly-coupled) dual Higgs picture worked.   
 
The checks made in this paper have been primarily aimed at ascertaining that one is indeed correctly  describing
 the infrared physics of these SCFT's of highest criticality, {\it  deformed by a small  $\mu \Phi^{2}$ perturbation}, in terms of the GST duals \cite{GST}.  
 Once such a test is done, one can safely discuss the infrared physics in the limit of singular  SCFT, directly,
%  \footnote{We verified that the low-energy GST description  (\ref{GSTD3}) and (\ref{noncollidr}) in the $SU(3)$, $SU(4)$ cases as well, does describe correctly the system; unfortunately it remains a non-local system.     }. 
 
Let us take the example of the theory  discussed in Section \ref{uspsection}.  
In the case of $USp(2N)$,  $N_{f}=4$ theory, the GST dual is 
 \beq       {\boxed 1} -   SU(2) - {\boxed 4}\;.
 \eeq
  The effect of $\mu \, \Phi^{2}$ deformation of this particular theory can then be analyzed straightforwardly  in the massless theory \cite{noi2}  by using the superpotential,  
\beq     \sqrt{2} \, Q_{0} A_{D} {\tilde Q}^{0} +  \sqrt{2} \, Q_{0} \phi {\tilde Q}^{0} + \sum_{i=1}^{4}  \sqrt{2} \, Q_{i} \phi {\tilde Q}^{i}  +  \mu A_{D} \Lambda  +  {\mu}\, \Tr \phi^{2}\;.
\eeq
The vacuum of this system was found in \cite{noi2}:
\beq  Q_{0} =  {\tilde Q}_{0}= \left(\begin{array}{c}   2^{-1/4}\sqrt{-\mu \Lambda}  \\ 0 \end{array}\right)
\eeq
\beq   \phi =0, \quad  A_{D}=0\;.  \label{sol2}
\eeq 
The contribution from $Q_{i}$'s must then cancel that of   $Q_{0}$ in  Eq.~(\ref{eq3}).   
By flavor rotation  the nonzero VEV can be attributed to $Q_{1}, {\tilde Q}^{1}$, i.e., either of the form
\beq     (Q_{1})^{1} =  ({\tilde Q}^{1})_{1} =2^{-1/4} \sqrt{\mu \Lambda} \;,    \qquad  Q_{i}={\tilde Q}_{i}=0, \quad i=2,3,4.    \label{sol3a}
\eeq
or  
\beq     (Q_{1})^{2} =  ({\tilde Q}^{1})_{2}   = 2^{-1/4}  \sqrt{-  \mu \Lambda} \;,    \qquad  Q_{i}={\tilde Q}_{i}=0, \quad i=2,3,4. \label{sol3b}
\eeq
 The $U(1)$ gauge symmetry is broken by the $Q_{0}$ condensation:  an ANO vortex is formed. As the gauge group of the underlying theory is  simply connected,
such a low-energy vortex must end. The quarks are  confined. The flavor symmetry breaking 
\beq 
  SO(8) \to   U(1) \times  SO(6) =   U(1) \times SU(4) = U(4),  \label{right}
\eeq
is induced by the condensation of  $Q_{1}$, which does not carry the $U(1)$ gauge charge.  The pattern of the symmetry breaking agrees with that found at large $\mu$  \cite{CKM}.

The vortex is made of the $Q_{0}$ field and the effective Abelian gauge
field.      The most interesting feature of this system is that there is no dynamical Abelianization, i.e., the effective low-energy gauge group is $SU(2)\times U(1)$.  
The confining string is unique and does not leads to the doubling of the meson spectrum. The global symmetry breaking of the low-energy effective theory is the 
right one (\ref{right}), but the vacuum is not color-flavor locked. The confining string is of Abelian type, and is not a non-Abelian vortex as the one appearing in an $r$ vacuum \cite{noi2}.  
These facts clearly  distinguish the confining system found here both from the standard Abelian dual superconductor type systems and from the 
non-Abelian dual Higgs system found in the $r$-vacua of SQCD.   The dynamical symmetry breaking and confinement are linked to each other (the former is induced by the $Q$ condensates, which in turn, is triggered by the $Q_{0}$ condensation which is the order parameter of confinement), but not described by one and the same condensate.  

The $SO(N)$ systems discussed  in Section \ref{SO(N)} present other examples of confining vacua, with similar properties. 

We conclude with a brief comment on the nature of the GST variables. The mass assignment such as in Eq.~(\ref{masses}) which reproduces correctly the flavor symmetry property of the underlying theory, is a clear sign of the magnetic monopole nature of the low-energy matter content. Their condensation therefore implies confinement of the color-electric charges. Nevertheless, the way they realize the dynamical flavor symmetry breaking and confinement appears to present various new features as compared to the 
straightforward dual superconductor picture of confinement, Abelian or non-Abelian, and seems to urge a better understanding of the new confinement phases.

\end{document}